\documentclass[12pt]{iopart}

\usepackage{graphicx}

\expandafter\let\csname equation*\endcsname\relax
\expandafter\let\csname endequation*\endcsname\relax
\usepackage{amsmath}

\usepackage{hyperref}
\usepackage{color}

\usepackage{iopams}

\begin{document}

\title[Sea-level rise and continuous adaptation]{Sea-level rise and continuous adaptation: residual damage rises faster than the protection costs}

\author{Diego Rybski$^{1,2}$, Boris F.\ Prahl$^1$, Markus Boettle$^1$, \& J\"urgen P.\ Kropp$^{1,3}$}

\address{
$^1$ Potsdam Institute for Climate Impact Research -- PIK, Member of Leibniz Association, P.O. Box 601203,
14412 Potsdam, Germany\\
$^2$ Department of Environmental Science Policy and Management, University of California Berkeley, 130 Mulford Hall \#3114, Berkeley, CA 94720, USA\\
$^3$ Institute for Environmental Science and Geography, University of Potsdam, 14476 Potsdam, Germany
}

\ead{ca-dr@rybski.de}

\vspace{10pt}
\begin{indented}
\item[]\today
\end{indented}

\begin{abstract}
Damage cost curves -- relating the typical damage of a natural hazard to its physical magnitude -- represent an indispensable ingredient necessary for climate change impact assessments.
Combining such curves with the occurrence probability of the considered natural hazard, expected damage and related risk can be estimated.
Here we study recently published city scale damage cost curves for coastal flooding and demonstrate which insights can be gained from the functions only.
Therefore, we include protection cost curves -- relating the typical investment costs necessary to protect a city against a natural hazard of certain magnitude -- which are analogous to and consistent with the above mentioned damage cost curves.
Specifically, we motivate log-logistic functions, which exhibit a power-law increase at the lower end, and fit them to the cost curves.
As expected, cities with large maximum potential loss (typically large cities) are also more costly to protect.
Moreover, we study the idealized case of continuous adaptation, i.e.\ increasing protection levels in the same pace as sea-level rise, and compare the associated costs with residual damage from extreme events exceeding the protection.
Based on the fitted exponents we find that in almost all cities the residual damage rises faster than the protection costs.
Raising coastal protection can lead to lull oneself in a deceptive safety.
\end{abstract}

\section*{Introduction}

\begin{figure}
\begin{center}
\includegraphics[width=0.85\textwidth]{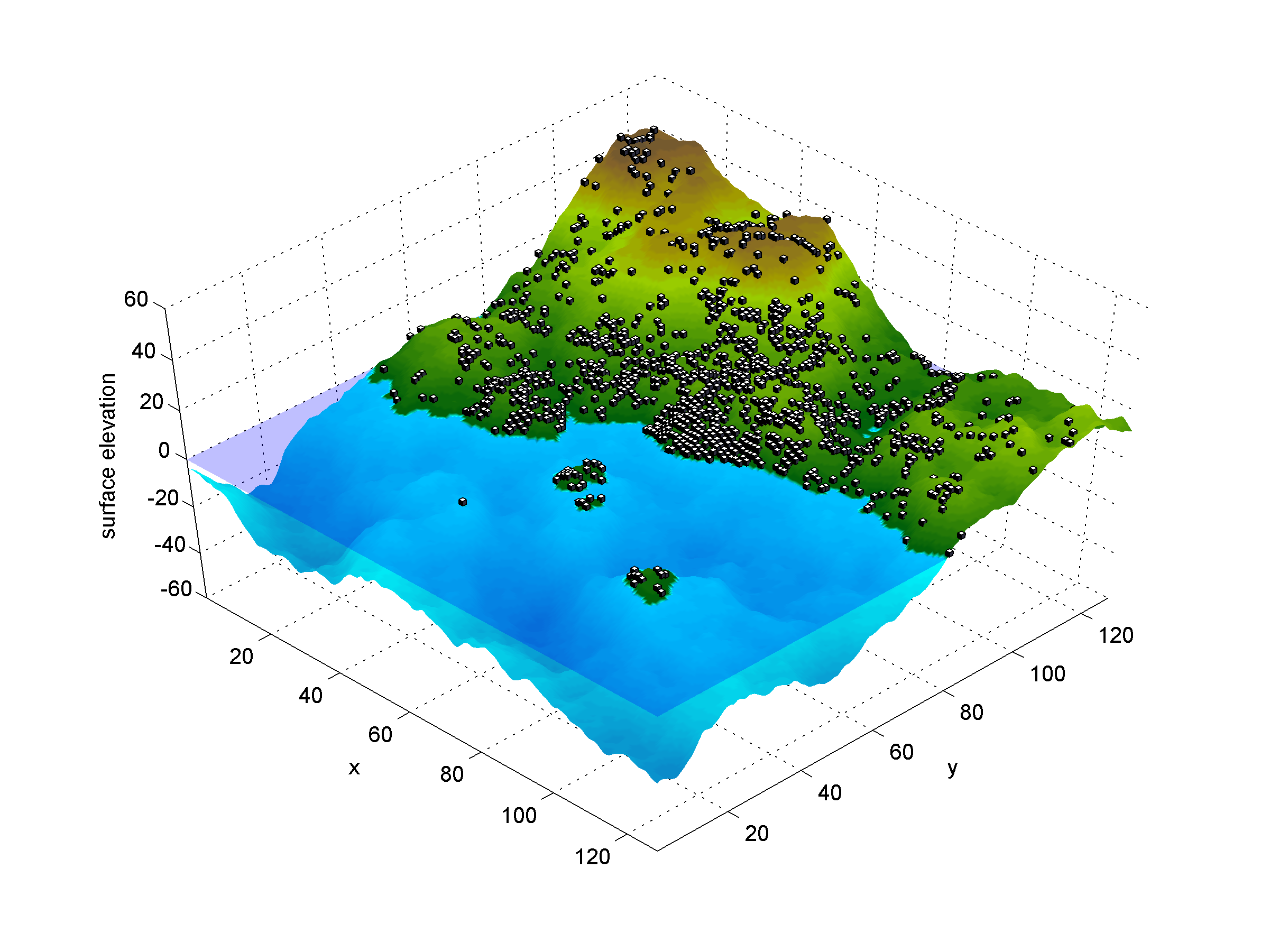}
\caption{Illustration of orography, sea-level, and urban areas (procedurally generated).}
\label{fig:fractalf}
\end{center}
\end{figure}

In most cases climate change impact and natural hazard assessments roughly consist of three components \cite{BoettleRK2016}.
First, the physical hazard, e.g.\ flood, storm, etc., exhibits a magnitude which is quantified e.g.\ by the maximum flood level or the maximum wind speed, respectively.
Since only the extreme magnitudes cause damage, their occurrence probability or annuality are considered.
Second, damage cost curves are used to estimate an expected damage associated to a hazard event of certain magnitude.
Here we refer to aggregated, city-wide cost curves, unless specified otherwise, see Fig.~\ref{fig:fractalf} for an illustration.
Cost curves are monotonously increasing since hazards of higher magnitude cause more damage.
Third, combining the hazard probability with the cost curves leads to the damage probability.
The full damage distribution \cite{AbadieGMM2019} function can be obtained from the hazard distribution function via cost curves (see e.g.\ \cite[Usage Notes]{PrahlBCKR2018}).

Although being at the core of damage assessments, cost curves play a marginal role in the corresponding publications.
Prahl~et~al.~\cite{PrahlBCKR2018} even claim that they are ``commonly regarded as by-products of impact assessments''.
One can distinguish empirical and synthetic damage relations \cite{MerzKST2010}.
The former are inferred by relating observed damage costs and associated hazard magnitudes.
The latter rely on standardized data bases or long-term surveys and involve simplifications and subsequent uncertainties \cite{MerzKST2010}.
Here we propose a third category, namely mathematical functions which can be fit to either category. E.g.\ power-law damage functions have been used for daily storm damage to residential buildings on the district level \cite{PrahlRKBH2012}. 
The advantage of such functional forms is that they can be studied mathematically \cite[e.g.]{BoettleRK2016}.

Here we consider damage functions as regressions to synthetic cost curves.
Specifically, we analyze damage cost curves for coastal flooding in European cities that have been compiled in a comparable manner \cite{PrahlBCKR2018}.
We find that log-logistic functions are adequate in most cases and motivate the shape of the function in the lower range.
Including protection cost curves that are defined on the same spatial units, allows us to explore similarities of damage and protection cost curves.
We find that costs for increasing coastal protection (marginal costs) can as well be described by log-logistic functions.
Comparing the fit parameters it turns out that the maximum potential loss and the maximum marginal protection costs correlate, which confirms the intuition that the same way as more damage can occur in larger cities also necessary investment costs to protect them are higher in larger cities.

With increasing sea-levels \cite{JevrejevaMG2010}, it can be expected that authorities regularly assess the coastal protection \cite{Dantzig1956}.
Comparing present and upcoming sea-levels, repeated raises of protection levels (design flood) might be necessary together with maintenance \cite{DawsonBWWHR2011,VousdoukasMHWMCF2020}.
Accordingly, during a century there might be individual or few augmenting modifications of the levees etc.
Here we approximate this step-wise increments by a continuous process where the protection level is assumed to increase with the same pace as sea-level rise \cite{HallBNPW2012}.
Under this scenario, we compare investment costs necessary along time to perform the raise of protection levels with the residual costs stemming from damage that occurs from rare surges that are beyond the protection levels (e.g.\ for Frechet extreme value characteristics such residual damage is unavoidable).
We obtain an analytic expression which involves one parameter from the damage function and another one from the protection function.

Using the values obtained from fitting the damage and protection cost curves we find that in 99\,\% of the considered cities, the residual damage rises faster than the costs of continuous adaptation.
In this scenario the number of events exceeding the protection level remains constant, but the damage caused by individual events increases.
This aspect adds to the problem of (possibly) false safety \cite{AnderssonSkoldTRLJJMG2015} behind protection measures.

\section*{Datasets}

We use the damage and protection cost curves provided by Prahl~et~al.~\cite{PrahlBCKR2018} for European coastal cities.
While originally cost curves for 600 cities were provided, we restrict ourselves to the top 100 largest  in area for which the fitting is more robust.

The damage cost curves consist of estimated direct monetary damage to the considered city by a hypothetical flood of maximum flood level between $0$\,m and $12$\,m in steps of $0.5$\,m.
The damage cost curves are based on static inundation (also known as flood-fill algorithm or bathtub model \cite{VousdoukasVMDGBBSF2016}) where all hydraulically connected areas are flooded according to the presumed flood height.

The protection cost curves provide -- analogous to the damage cost curves -- the costs necessary to protect the considered city against a hypothetical flood of maximum flood level (design height) between $0$\,m and $12$\,m in steps of $0.5$\,m.
They consist of estimated costs to construct levees following the \emph{urban protection course}, i.e.~the required location of protection measures.
Maintenance is not included.
The estimates refer to the situation without protection measure but respective information can be taken into account \cite{PrahlBCKR2018} if available.
Low existing protection measures can be neglected for marginal costs of higher protection levels.

\section*{Results}

\subsection*{Regressing damage and protection cost curves}
\begin{figure}
\begin{center}
\includegraphics[width=0.85\textwidth]{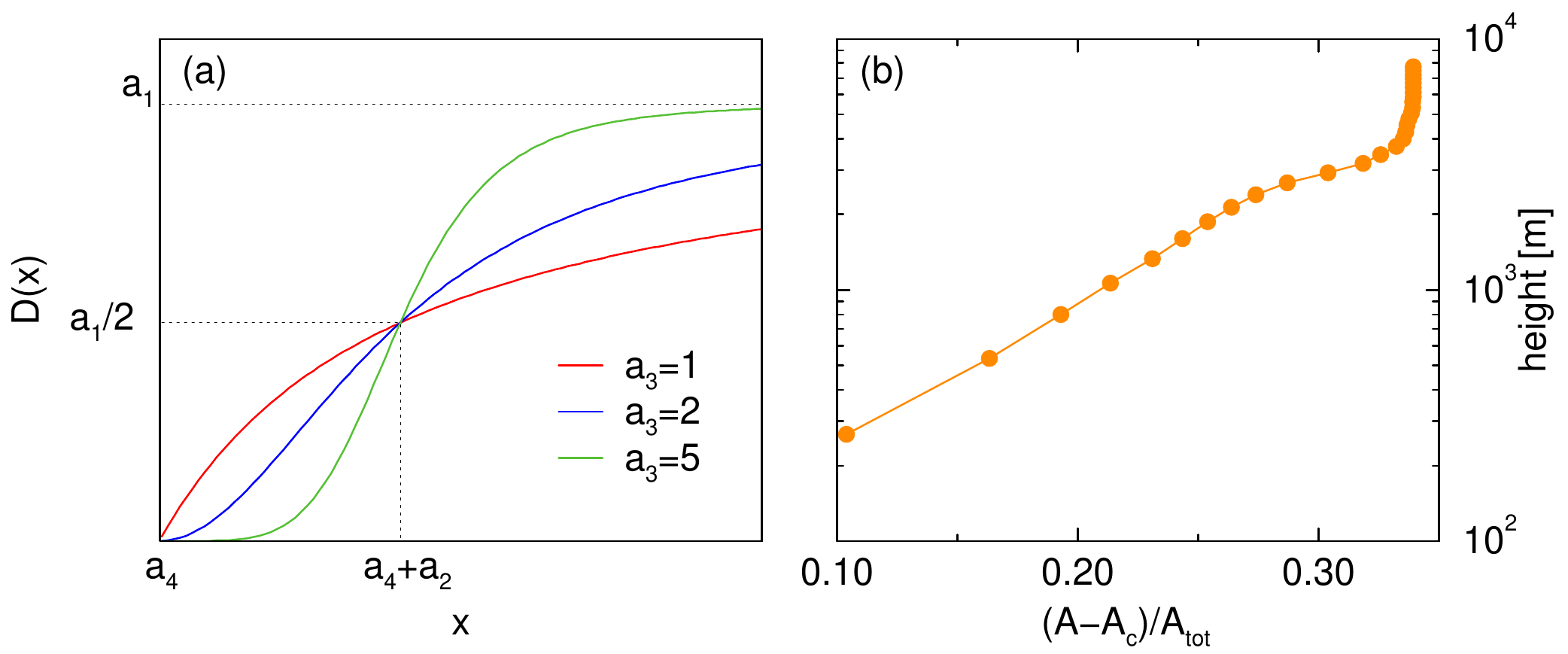}
\caption{Illustration of log-logistic function and global ocean surface.
(a) Illustration of Eq.~(\ref{eq:Dx}) for three different values of $a_3$.
(b) Relation between flooded area and height (data from \cite{SaberiAA2013}).
$A_\textrm{c}$ is the present sea-level and $A_\textrm{tot}$ is the total surface of the globe.
There is an approximate exponential relation below $2,000$\,m, which in the lower range can be approximated by a linear form.
}
\label{fig:logSaberi}
\end{center}
\end{figure}

\begin{figure}
\begin{center}
\includegraphics[width=0.85\textwidth]{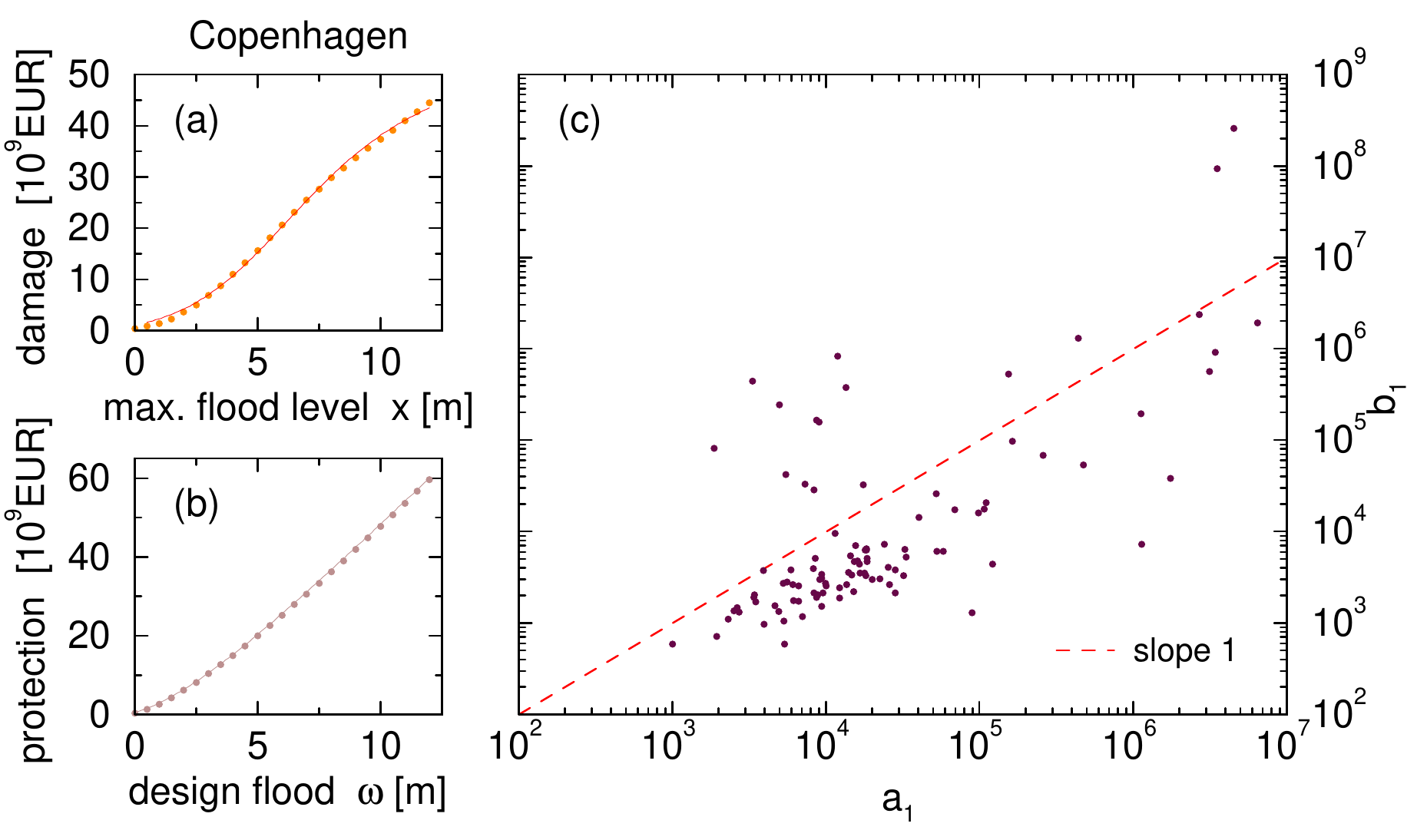}
\caption{Copenhagen example and correlations between parameters.
Panels~(a) and~(b) display the damage and protection cost curves \cite{PrahlBCKR2018} (dots) together with regressions according to Eq.~(\ref{eq:Dx}) and~(\ref{eq:Pomega}) (solid lines), respectively.
In panel~(c) the parameters $a_1$ and $b_1$ obtained for the top~100 largest cities are plotted against each other.
The dashed straight line has slope~$1$.
}
\label{fig:copenhagen}
\end{center}
\end{figure}

Damage cost curves provide the typical damage that can be expected from a natural hazard (here storm surges) of certain magnitude (here maximum flood level).
We fit the following log-logistic model \cite{RybskiDK2020} (which is also known as Hill function) to the damage costs \cite{PrahlBCKR2018} $D$ as a function of maximum flood level $x$
\begin{equation}
D(x) = \frac{a_1}{1+(\frac{x-a_4}{a_2})^{-a_3}}, \qquad x>a_4
\label{eq:Dx}
\, 
\end{equation}
where $a_1$\dots{}$a_3$ are fitting parameters and $a_4$ is the lowest urban elevation within the considered city. 
This form assumes, that there is no natural or artificial flood protection in place and needs to be adjusted by setting $D(x)=0$ for $x$ below an existing protection level $\omega$ ($>a_4$).
The parameters can be interpreted as $a_1$ maximum potential loss, $a_2$ determining the inflection, and $a_3$ steepness, see Fig.~\ref{fig:logSaberi}(a).
For $a_2=1, a_3=1, a_4=0$ Eq.~(\ref{eq:Dx}) becomes $D(x)/a_1=\frac{x}{x+1}$ as employed in \cite{HinkelLVPNTMFIL2014}.
The inflection point of $D(x)$ is located at $x^*=a_2+a_4$, $D(x^*)=\frac{a_1}{2}$.
The parameters $a_2$ and $a_4$ are in meters, the unit of $a_1$ is the considered currency, and $a_3$ carries no unit.
For small flood levels, Eq.~(\ref{eq:Dx}) provides the approximation 
\begin{equation}
D(x) \sim x^{a_3} \label{eq:Dxapprox}
\, .
\end{equation}
The power-law form has e.g.\ been used to assess coastal flood damage and protection scenarios based on extreme value theory \cite{BoettleRK2016,BoettleRK2013}.
This approximation is justified as typically the inflection point of the sigmoid curve is clearly beyond the relevant flood magnitudes, e.g.\ $x^*\approx 6$\,m for Copenhagen in Fig.~\ref{fig:copenhagen}(a), see \cite{PrahlBCKR2018} for further examples.

We employ an analogous functional form for the protect costs \cite{PrahlBCKR2018} $P$ as a function of the protection level (design height) $\omega$.
In \ref{sec:protfunc} we motivate that here a cumulative form is necessary,
\begin{equation}
P(\omega) = \int_{\omega_0}^\omega f(z)\,\textrm{d}z, \qquad f(z) =
\frac{b_1}{1+(\frac{z-b_4}{b_2})^{-b_3}}, \qquad \omega\geq\omega_0\label{eq:Pomega}
\,
\end{equation}
where the fitting parameters $b_1$\dots{}$b_3$ are analogous to the ones in Eq.~(\ref{eq:Dx}), $b_4<z$ is the lowest elevation along the protection course and $\omega_0$ represents the existing protection level (the case without existing flood protection corresponds to $\omega_0=b_4$). The function $f(z)$ can be understood as marginal costs, i.e.~the costs to raise the protection level from $z$ by one unit.
For small protection levels and $\omega_0\approx b_4$, Eq.~(\ref{eq:Pomega}) approximately follows
\begin{equation}
P(\omega) \sim \omega^{b_3+1} \, .\label{eq:Pwapprox}
\end{equation}
This power-law form is analogous to Eq.~(\ref{eq:Dxapprox}).

In Fig.~\ref{fig:copenhagen} we show an example of the fitted functions.
For Copenhagen, the damage costs, Fig.~\ref{fig:copenhagen}(a), show a sigmoidal shape with a turning point towards saturation.
The three regression parameters provide sufficient flexibility to fit the values of the damage curve.
The protection costs, Fig.~\ref{fig:copenhagen}(b), exhibit a weak curvature without turning point.
Due to the cumulative character (increasing protection level can build on existing levee), and the characterization via the integral over the log-logistic function in Eq.~(\ref{eq:Pomega}), the damage costs do not reach saturation, but instead approach a linear asymptote.
This is plausible, as with increasing protection levels the levee height continues to increase and so do the protection costs (assuming linear unit costs \cite{LenkRHDK2017}).
As long as the protection course is not closed around the considered city, $P(\omega)$ increases steeper than linearly and the approximation Eq.~(\ref{eq:Pwapprox}) is justified, see Fig.~\ref{fig:copenhagen}(b) and \cite{PrahlBCKR2018} for further examples.

Comparing the obtained fitting parameters $a_1$\dots{}$a_3$, $b_1$\dots{}$b_3$ for the top 100 largest  considered by Prahl~et~el.~\cite{PrahlBCKR2018} we find correlations between $a_1$ and $b_1$ as displayed in Fig.~\ref{fig:copenhagen}(c).
This means, in cities where the potential damage is high, also the protection costs tend to be high.
Expectedly, it is the large cities where both damage and protection costs can be high.

\subsection*{Comparing residual damage and protection costs (continuous adaptation scenario)}

We want to investigate the scenario in which the protection is continuously adjusted to sea-level rise \cite{HallBNPW2012}.
Hence we assume adaptation \cite{AdgerAT2005,RosenzweigSHM2010,DowBPKMS2013,NalauPM2015} in order to maintain flood protection against a predefined return period, e.g.~the 1000-year event. 
We also make the approximation that sea level rise uniformly shifts the distribution of flood heights.
Extreme flood heights are described by a Generalized Pareto Distribution with location parameter $\mu$.

For continuous adaptation, we asymptotically set the protection level $\omega$ equal to the location parameter, i.e.\ $\omega=\mu$.
In \ref{app:contadapt} we show that given a power-law damage function Eq.~(\ref{eq:Dxapprox}), the residual expected annual damage, i.e.\ the damage of those flood events exceeding the protection \cite{PrahlR2019}, is then given by $\textrm{E}_D(\mu) \sim \mu^{a_3}$.
If the distribution of extremes is simply moved with the sea-level rise (i.e.\ altered location but constant shape and scale), then the number of floods which exceed the protection level will be the same (if the design flood is continuously adapted to sea-level rise), but the damage can be different and increases with sea-level rise (the higher the protection, the more destructive can an exceeding event be).
In other words, under continuous adaptation the exceedance probability of extreme events remains constant but the risk (as the product of probability times loss) continues to increase.

Annual cost to raise the design flood level adjusting to sea-level rise $\textrm{E}_P(\mu)$ (in the following \emph{annual protection cost} for short) is estimated as $\textrm{d}P(\mu)/\textrm{d}\mu \cdot\Delta \mu$. 
Due to the strong inertia of sea level rise, the annual increments $\Delta\mu$ can be considered as approximately constant. 
Hence, annual protection costs are given approximately by $\textrm{E}_P(\mu)\sim\mu^{b_3}$ (it is worth noting that these costs do not include maintenance).

In order to compare the protection costs with the residual damage, we consider the fraction
\begin{align}
\frac{\textrm{E}_P(\mu)}{\textrm{E}_D(\mu)}&\sim\mu^{b_3-a_3}
\label{eq:ratio}
\, .
\end{align}
One can see that if $b_3>a_3$, the annual protection costs rise faster than the annual residual damage and if $b_3<a_3$, the annual protection costs rise slower than the annual residual damage.
Please note that Eq.~(\ref{eq:ratio}) is independent of already installed protection measures as long as Eqs.~(\ref{eq:Dxapprox}) and~(\ref{eq:Pwapprox}) are valid approximations, but sufficiently large $\mu$ are required \cite{BoettleRK2016}.

\begin{figure}
\begin{center}
\includegraphics[width=0.75\textwidth]{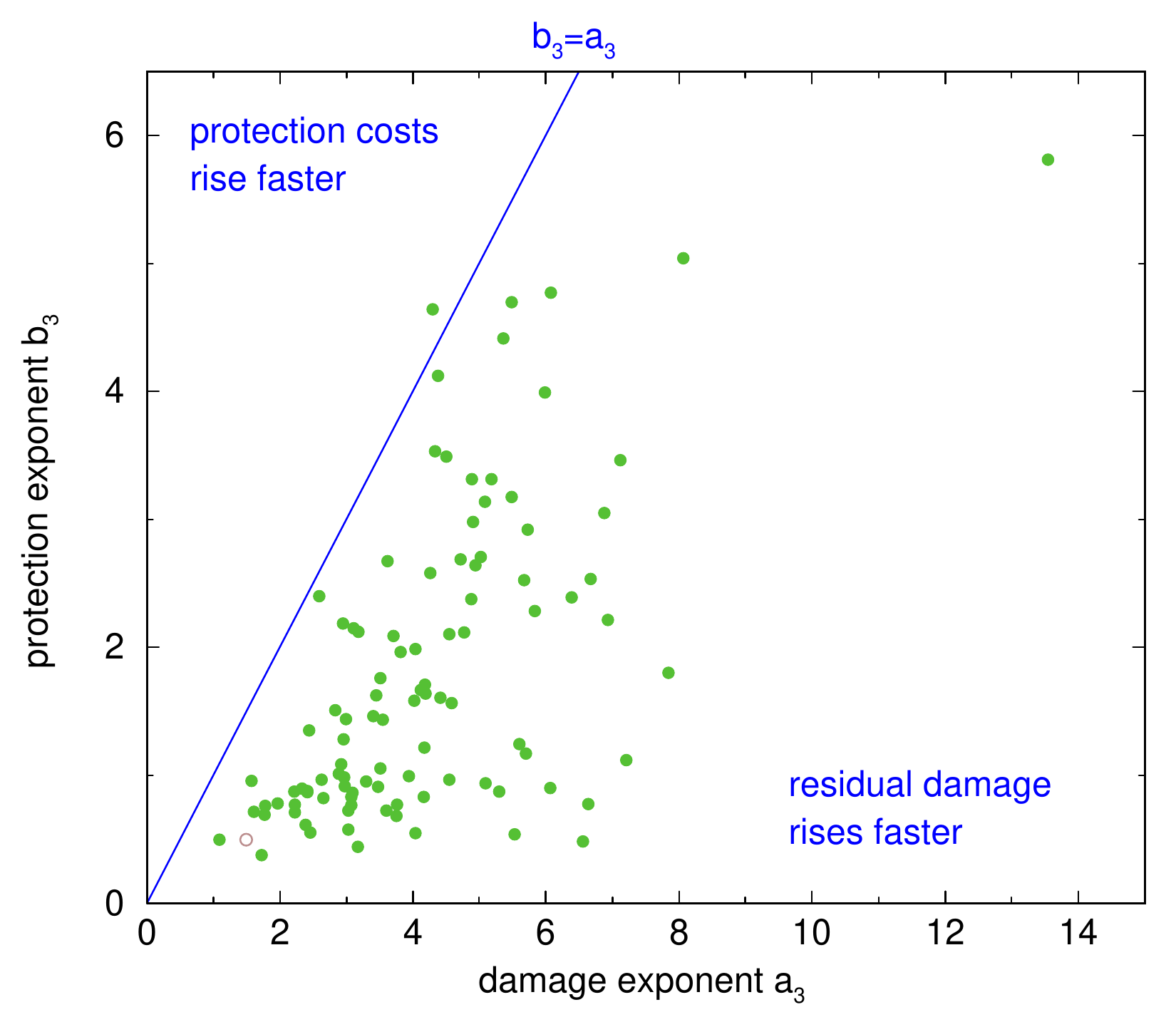}
\caption{Relating protection and damage cost exponents. 
The exponent $b_3$ from Eq.~(\ref{eq:Pomega}) is plotted vs.\ the exponent $a_3$ from Eq.~(\ref{eq:Dx}).
The blue line is given by $b_3=a_3$.
Each green dot represents the values of a city.
In cities above (1 out of 100), the protection costs rise faster, in cities below (99 out of 100), the residual damage rises faster.
The brown open symbol represents the values of a circular city on an inclined plane ($a_3=3/2$, assuming $\alpha_a,\alpha_m=1$) where the protection costs are given by the length of flooded city border, i.e.\ the arc which scales as $b_3= 1/2$.
}
\label{fig:expos}
\end{center}
\end{figure}

In Fig.~\ref{fig:expos} we plot the estimates of both exponents, $b_3$ and $a_3$, against each other together with a line indicating the case $b_3=a_3$.
It can be seen that $a_3>b_3$ in most cases, i.e.\ annual residual damage rises faster than the annual protection costs in most of the 100 considered cities.
Equation~(\ref{eq:ratio}) does not allow any conclusion in absolute terms.
However, the finding $a_3>b_3$ for most cities, can also be rephrased in the following way. 
If the ratio of protection and damage costs is to be kept constant, then the protection level needs to be increased at a higher pace than sea-level rise.

\section*{Discussion \& Outlook}

In summary, we propose and motivate log-logistic damage and protection cost functions.
Such functional forms can be fitted to empirical and synthetic damage cost curves.
We use them to characterize damage and protection at the 100 largest coastal cities in Europe and find (i) that cities with higher potential loss are also more expensive to protect and (ii) that residual damage rises faster than protection costs when protection levels are increased with sea-level rise.

Our work involves a set of uncertainties.
The regressions depend on the accuracy of the cost curves. 
On the one hand, bottom-up case studies might provide better estimates.
But on the other hand, the systematic and comparable approach \cite{PrahlBCKR2018} allows the comparison of a large number of cities as done here.
Depending on how pronounced the asymptotic behavior is represented in the data, the parameters $a_1$ and $b_1$ might be affected by more or less uncertainty.
Last but not least, the cost curves of small cities are more noisy and exhibit erratic bumps so that here we restrict ourselves to the largest cities.

The exponent $a_3$ is a combination of various factors.
It consist of the relations how the flooded terrain increases with flood height \cite{MerzET2009}, how the flooded \emph{urban} area increases with growing flooded terrain, and how the local damage increases with the local flood level at an urban site (e.g.\ pixel or building).
\begin{itemize}
\item Orography.
The flooded terrain as a function of the flood height follows $A(x)\sim x^{\alpha_a}$, where $A$ is the area of the flooded terrain and $\alpha_a$ is the exponent relating it to the flood height.
From \cite{SaberiAA2013} we infer $x\simeq\exp{(a(A-A_\textrm{c})+b)}$ for $0<x<2000$, were $a$ and $b$ are parameters.
For $0\leftarrow (A-A_\textrm{c})$ the exponential function can be approximated by $x\approx 1+a(A-A_\textrm{c})+b$, i.e.\ a linear relation ($\alpha_a\approx 1$) on the global average, see Fig.~\ref{fig:logSaberi}(b).
However, there are also regional differences, e.g.\ $\alpha_a\approx 2$ reported for the Kalundborg case study \cite{BoettleKRRRW2011}.
\item Urban Morphology.
The flooded \emph{urban} area as a function of the flooded terrain follows $U(A)\sim A^{\alpha_u}$, where $\alpha_u$ is another exponent.
If we consider the idealized case of a city with circle shape on an inclined plane then the flooded urban area is given by the area of the circular segment $U(A)=\arccos(1-A)-(1-A)\sqrt{2A-A^2}$, which for $0\leftarrow A$ approximately behaves as $U(A)\sim A^{3/2}$, i.e.\ $\alpha_u=3/2$.
Different exponents are possible for more complex morphologies.
\item Micro Damage.
The damage of the micro-elements as a function of local flood levels follows $g(x-\lambda)\sim (x-\lambda)^{\alpha_m}$ for $0<g(x-\lambda)<g_\textrm{max}$ (see \ref{sec:protfunc} and Fig.~\ref{fig:fractalf}).
For the Kalundborg case study \cite{BoettleKRRRW2011} linear, square-root, and quadratic forms have been exemplified, i.e.\ exponents $\alpha_m=1,1/2,2$, respectively.
\end{itemize}
Combining these three factors leads to $a_3\approx \alpha_a\,\alpha_u\,\alpha_m$.
As it is a product, large $a_3$ can results if each exponent is larger than 1, e.g.\ $(1.5)^3=27/8\approx 3.4$.

As an outlook it could be interesting to rephrase the research question and ask, how the protection levels need to be increased so that the residual damage remains constant with sea-level rise.
We are confident that it is possible to address this question but it requires more efforts along the lines of \cite{BoettleRK2016,BoettleRK2013}.
More on the data side, it could be helpful, to infer the parameters $a_i$ and $b_i$ directly from features of the terrain and the urban morphology, e.g.\ roughness etc.
A better understanding of the genesis of the parameters would permit to estimate them without work intensive calculations \cite{PrahlBCKR2018}.

\appendix

\section{Protection Functions}
\label{sec:protfunc}
Drawing an analogy between damage and protection cost curves, we want to deduce a general mathematical form for protection cost curves. Specifically, we show that for a constant unit cost function the protection cost curves are given by an integral over the cumulative frequency distribution of protection segments. 

For a group of similar elements, e.g.\ a city comprised of numerous buildings, we can formulate a damage cost curve as the convolution of a micro-scale damage function, $d(\cdot)$, and the probability density, $f_\Lambda(\lambda)$, of the hazard thresholds of the individual elements \cite{PrahlRBK2016}:
\begin{eqnarray}
D(x) &=& (f_\Lambda * d)(x)\nonumber \\
	&=& \int_0^x f_\Lambda(\lambda)d(x-\lambda)\,\textrm{d}\lambda
\, .
\end{eqnarray}
By analogy, we can formulate a corresponding expression for the protection cost curve
\begin{equation}
P(\omega) = \int^\omega_{\omega_0}{f'(z)p(\omega-z)\,\textrm{d}z}
\, ,
\label{eq:A2}
\end{equation}
where $p(\cdot)$ is the cost function for the installation of levees and $f'(z)$ denotes the frequency of protection segments with foundation at height $z$.
The integral's lower bound $\omega_0$ denotes the minimum flood height at which protection is required.

The unit costs for raising levees are approximately independent of the height of the levees and can be assumed constant \cite{JonkmanHNKL2013,LenkRHDK2017}, hence $p(\omega-z)\sim (\omega-z)$. We use the constant unit cost assumption and integration by parts to expand Eq.~(\ref{eq:A2}):
\begin{align}
P(\omega) &\sim \int^\omega_{\omega_0}{f'(x)(\omega-x)\,\textrm{d}z}\nonumber\\[2ex]
&= \omega\int^\omega_{\omega_0}{f'(z)\,\textrm{d}z} - [zf(z)]^\omega_0 + \int^\omega_{\omega_0}{f(z)\,\textrm{d}z}\nonumber\\[2ex]
&= \omega[f(z)]^\omega_{\omega_0} - [zf(z)]^\omega_{\omega_0} + [F(z)]^\omega_{\omega_0}\nonumber\\[2ex]
&= F(\omega) - F({\omega_0}) - \omega f({\omega_0})
\, .
\label{eq:4}
\end{align}
Per definition $f(\omega_0)=0$ and $F(\omega_0)=0$. Hence we obtain our final result
\begin{equation}
P(\omega)\sim F(\omega).
\end{equation}

Since $f'(z)$ represents the frequency distribution of the foundation height of different segments of the protection course, i.e.\ typically a bell-shaped curve, $f(z)$ is the corresponding cumulative frequency distribution, which generally takes a sigmoidal form.
Moreover, it follows that $F(\omega)$, the integral of $f(z)$, increases linearly for asymptotic large $\omega$.

\section{Expected annual damage under continuous Adaptation}
\label{app:contadapt}

Employing a point process and the \emph{Generalized Pareto Distribution} to model the occurrence and magnitude of flood events as proposed by Boettle~et~al.~\cite{BoettleRK2016}, we want to deduce, that in the considered scenario of continuous adaptation, the expected residual damage $\textrm{E}_D$ becomes roughly proportional to the damage Function $D$.

Given a power law damage function $D(x)\sim x^{a_3}$ (see Eq.~\ref{eq:Dxapprox}) and a continuous adjustment of the protection level to the rise of the mean sea level (i.e.\,$\mu=\omega$), it holds~\cite{BoettleRK2016}:
\begin{enumerate}
\item The expected number of floodings $\Lambda$ is constant.

\item The expected damage caused by a singular flooding equals
\begin{eqnarray}
\textrm{E}_{D_i} &=& \int_{\omega}^{\infty}D(x)p(x;\xi,\sigma,\omega)\textrm{d}x\\
&=& \int_{0}^{\infty}D(x+\omega)p(x;\xi,\sigma,0)\textrm{d}x \, ,
\end{eqnarray}
where $\mu, \sigma$ and $\xi$ are the parameters of the \emph{Generalized Pareto Distribution} $p$.
\item The expected annual damage is
\begin{eqnarray}
\textrm{E}_D=\Lambda\cdot \textrm{E}_{D_i} \, .
\end{eqnarray}
\end{enumerate}

For large values of $\omega$, we can conclude that $\textrm{E}_{D_i}(\omega) \sim D(\omega)$ and since the average number of flood events is constant we find the same proportionality for the annual damage
\begin{eqnarray}
\textrm{E}_D(\omega) &=& \Lambda\cdot\textrm{E}_{D_i}(\omega)\\
&\sim& D(\omega) \, .
\end{eqnarray}

\section*{Acknowledgements}
We thank A.A.~Saberi for providing the data displayed in Fig.~\ref{fig:logSaberi}(b).
The research leading to these results has received funding from the European Community's Seventh Framework Programme under Grant Agreement No.~308497 (Project RAMSES).
D.\ Rybski thanks the Alexander von Humboldt Foundation for financial support under the Feodor Lynen Fellowship.

\section*{References}

\bibliographystyle{iopart-num}
\bibliography{arxiv}

\providecommand{\newblock}{}
\begin{thebibliography}{10}
\expandafter\ifx\csname url\endcsname\relax
  \def\url#1{{\tt #1}}\fi
\expandafter\ifx\csname urlprefix\endcsname\relax\def\urlprefix{URL }\fi
\providecommand{\eprint}[2][]{\url{#2}}

\bibitem{BoettleRK2016}
Boettle M, Rybski D and Kropp J~P 2016 {\em Nat. Hazards Earth Syst. Sci.\/}
  {\bf 16} 559--576

\bibitem{AbadieGMM2019}
Abadie L~M, Galarraga I, Markandya A and de~Murieta E~S 2019 {\em Environ. Res.
  Lett.\/} {\bf 14} 064021

\bibitem{PrahlBCKR2018}
Prahl B~F, Boettle M, Costa L, Kropp J~P and Rybski D 2018 {\em Sci. Data\/}
  {\bf 5} 180034

\bibitem{MerzKST2010}
Merz B, Kreibich H, Schwarze R and Thieken A 2010 {\em Nat. Hazards Earth Syst.
  Sci.\/} {\bf 10} 1697--1724

\bibitem{PrahlRKBH2012}
Prahl B~F, Rybski D, Kropp J~P, Burghoff O and Held H 2012 {\em Geophys. Res.
  Lett.\/} {\bf 39} L06806

\bibitem{JevrejevaMG2010}
Jevrejeva S, Moore J~C and Grinsted A 2010 {\em Geophys. Res. Lett.\/} {\bf 37}
  L07703

\bibitem{Dantzig1956}
van Dantzig D 1956 {\em Econometrica\/} {\bf 24} 276--287

\bibitem{DawsonBWWHR2011}
Dawson R~J, Ball T, Werritty J, Werritty A, Hall J~W and Roche N 2011 {\em
  Global Environ. Change\/} {\bf 21} 628–--646

\bibitem{VousdoukasMHWMCF2020}
Vousdoukas M~I, Mentaschi L, Hinkel J, Ward P~J, Mongelli I, Ciscar J~C and
  Feyen L 2020 {\em Nature Com.\/} {\bf 11} 1--11

\bibitem{HallBNPW2012}
Hall J~W, Brown S, Nicholls R~J, Pidgeon N~F and Watson R~T 2012 {\em Nat.
  Clim. Change\/} {\bf 2} 833--834

\bibitem{AnderssonSkoldTRLJJMG2015}
Andersson-Sk{\"o}ld Y, Thorsson S, Rayner D, Lindberg F, Janh{\"a}ll S, Jonsson
  A, Moback U, Bergman R and Granberg M 2015 {\em Climate Risk Management\/}
  {\bf 7} 31--50

\bibitem{VousdoukasVMDGBBSF2016}
Vousdoukas M~I, Voukouvalas E, Mentaschi L, Dottori F, Giardino A, Bouziotas D,
  Bianchi A, Salamon P and Feyen L 2016 {\em Nat. Hazards Earth Syst. Sci.\/}
  {\bf 16} 1841--1853

\bibitem{RybskiDK2020}
Rybski D, Dawson R~J and Kropp J~P 2020 {\em Nat. Hazards Rev.\/} {\bf 21}
  06019003

\bibitem{HinkelLVPNTMFIL2014}
Hinkel J, Lincke D, Vafeidis A~T, Perrette M, Nicholls R~J, Tol R~S~J, Marzeion
  B, Fettweis X, Ionescu C and Levermann A 2014 {\em Proc. Natl. Acad. Sci. U.
  S. A.\/} {\bf 111} 3292--3297

\bibitem{BoettleRK2013}
Boettle M, Rybski D and Kropp J~P 2013 {\em Water Resour. Res.\/} {\bf 49}
  1199--1210

\bibitem{SaberiAA2013}
Saberi A~A 2013 {\em Phys. Rev. Lett.\/} {\bf 110} 178501

\bibitem{LenkRHDK2017}
Lenk S, Rybski D, Heidrich O, Dawson R~J and Kropp J~P 2017 {\em Nat. Hazards
  Earth Syst. Sci.\/} {\bf 17} 765--779

\bibitem{AdgerAT2005}
Adger W~N, Arnell N~W and Tompkins E~L 2005 {\em Global Environ. Chang.\/} {\bf
  15} 77--86

\bibitem{RosenzweigSHM2010}
Rosenzweig C, Solecki W, Hammer S~A and Mehrotra S 2010 {\em Nature\/} {\bf
  467} 909--911

\bibitem{DowBPKMS2013}
Dow K, Berkhout F, Preston B~L, Klein R~J~T, Midgley G and Shaw M~R 2013 {\em
  Nat. Clim. Change\/} {\bf 3} 305--307

\bibitem{NalauPM2015}
Nalau J, Preston B~L and Maloney M~C 2015 {\em Environ. Sci. Policy\/} {\bf 48}
  89--98

\bibitem{PrahlR2019}
Prahl B~F and Rybski D 2019 Climate change adaptation: Reconciling optimization
  and amortization {\em The Cities and Climate Change Complex -- Proceedings of
  the Cities and Climate Change Conference Held in Potsdam 2017\/} ed Kropp J~P
  and Rybski D (New York: Springer-Verlag)

\bibitem{MerzET2009}
Merz B, Elmer F and Thieken A~H 2009 {\em Nat. Hazards Earth Syst. Sci.\/} {\bf
  9} 1033--1046

\bibitem{BoettleKRRRW2011}
Boettle M, Kropp J~P, Reiber L, Roithmeier O, Rybski D and Walther C 2011 {\em
  Nat. Hazards Earth Syst. Sci.\/} {\bf 11} 3327--3334

\bibitem{PrahlRBK2016}
Prahl B~F, Rybski D, Boettle M and Kropp J~P 2016 {\em Nat. Hazards Earth Syst.
  Sci.\/} {\bf 16} 1189--1203

\bibitem{JonkmanHNKL2013}
Jonkman S~N, Hillen M~M, Nicholls R~J, Kanning W and van {L}edden M 2013 {\em
  J. Coastal Res.\/} {\bf 29} 1212--1226

\end{thebibliography}

\end{document}